\documentclass[11pt, letterpaper]{article}
\usepackage[utf8]{inputenc}
\usepackage[margin=1.5cm]{geometry}
\usepackage{titlesec}
\usepackage{tabu}
\usepackage{enumitem}
\usepackage{amssymb}
\usepackage{xcolor}
\usepackage{todonotes}
\newlist{selectlist}{itemize}{2}
\setlist[selectlist]{label=$\square$,leftmargin=*,noitemsep,topsep=0pt}
\usepackage{biblatex} %Imports biblatex package
\usepackage{lmodern}

\usepackage{hyperref}
\usepackage{cleveref}
\usepackage{eurosym}
\hypersetup{
    colorlinks=true,
    linkcolor=blue,
    filecolor=magenta,      
    urlcolor=blue,
}
 
\titleformat{\section}[block]{\bfseries}{\thesection.}{0.5em}{} 
\titleformat{\subsection}[block]{}{\thesubsection}{0.5em}{}
\titleformat{\subsubsection}[block]{\itshape}{}{0em}{}
\titlespacing*{\subsubsection}{0pt}{1em}{0.3em}
\begin{document}
% Create the title block
\begin{flushleft}

\setlength{\parindent}{0pt}
\setlength{\parskip}{10pt}

%Insert title
\textbf{A Flexible Raspberry Pi-Based Data Logger Platform for Modbus Sensors with Ansible Deployment}\\ 
%Insert Authors
\textbf{Authors}:\\
Leon Keim*,
Steffen Hägele,
Vivien Langhans,
Holger Class

%Insert Affiliations
\textbf{Affiliations}\\
Institute for Modeling Hydraulic and Environmental Systems, University of Stuttgart, 70569 Stuttgart, Germany

%Insert Contact Email
\textbf{Corresponding author's email address}\\ 
leon.keim@iws.uni-stuttgart.de

%Insert Abstract
\textbf{Abstract}\\

This article presents LibrePiLogger, an open-source data logging platform based on the Raspberry Pi for environmental monitoring using Modbus sensors over RS-485. The system combines the AtmosPyre Python library for sensor communication with Ansible-based deployment automation, allowing researchers to deploy sensor networks by editing a single YAML inventory file. Two hardware configurations are described: a minimal setup using a Raspberry Pi Zero with an RS-485 HAT, and a maximal setup using a Raspberry Pi 4 with a USB-to-RS-485 converter. Currently implemented sensors include the Vaisala GMP252 for CO$_2$ and the RadonTech AlphaTRACER for $^{222}$Rn, with new sensors requiring approximately 100 lines of Python following a provided driver template. Data is logged to timestamped CSV files with JSON metadata. The system has been deployed for continuous CO$_2$ and $^{222}$Rn monitoring in a karst environment since spring 2025 and remains in active operation, demonstrating reliable long-term performance. All hardware designs, software, and deployment scripts are released under the GNU General Public License v3.0. Total hardware costs range from 54 to 63\euro{} (excluding housing), depending on the configuration.

%Insert Keywords
\newpage
\textbf{Specifications table}\\
\vskip 0.2cm
\tabulinesep=1ex
\begin{tabu} to \linewidth {|X|X[3,l]|}
\hline  \textbf{Hardware name} & LibrePiLogger
  \\
  \hline \textbf{Subject area} &
  \vskip 0.1cm
  \begin{itemize}[noitemsep, topsep=0pt]
  \item Environmental, planetary and agricultural sciences
  \end{itemize}
  \\
  \hline \textbf{Hardware type} &
  \vskip 0.1cm  
  \begin{itemize}[noitemsep, topsep=0pt]
  \item Field measurements and sensors
  \item Electrical engineering and computer science
  \end{itemize}
  \\ 
\hline \textbf{Closest commercial analog} &
  \begin{itemize}
      \item Radon Scout, AlphaE or AlphaGUARD for radon monitoring
      \item Vaisala handheld CO$_2$ meter
  \end{itemize}
  \\
\hline \textbf{Open source license} &
GNU General Public License v3.0 or later
  \\
\hline \textbf{Cost of hardware} &
  54--63\euro{} (excluding housing)
  \\
\hline \textbf{Source file repository} & 
  \begin{itemize}
        \item Design Files: \url{https://doi.org/10.5281/zenodo.20117804}
      \item AtmosPyre Software:
      \begin{itemize}
          \item DOI: \url{https://doi.org/10.18419/DARUS-5531}
          \item Zenodo: \url{https://zenodo.org/records/20118423}
          \item PyPI: \url{https://pypi.org/project/atmospyre/}
          \item Git: \url{https://git.iws.uni-stuttgart.de/measurements/atmospyre}
      \end{itemize}
      \item Deployment Scripts: 
      \begin{itemize}
        \item DOI: \url{https://doi.org/10.5281/zenodo.20034879}
        \item Git: \url{https://git.iws.uni-stuttgart.de/measurements/senso_pi}
      \end{itemize}
  \end{itemize}
\\\hline
\end{tabu}
\end{flushleft}

\newpage
\section{Hardware in context}

Long-term monitoring of gases such as CO$_2$ and $^{222}$Rn in epigenic karst systems is essential for understanding site-specific transport dynamics, where dilution and accumulation of gases are governed by ventilation patterns and source terms; source strength is seasonally varying for CO$_2$ but relatively constant for $^{222}$Rn. In many karst systems, much of the CO$_2$ originates from microbial activity in the vadose zone, where multiphase gas--water transport takes place. CO$_2$ dynamics, in particular, drive limestone and dolomite dissolution during karstification and strongly influence CO$_2$ uptake at the air--water interface of stagnant water bodies through density-driven convective dissolution rather than turbulent mixing \cite{class2020, class2023, keim2025}.\\

Continuous monitoring in remote, subterranean environments requires a robust, adaptable, and cost-efficient system capable of autonomous data acquisition, storage, and optional remote transmission. Numerous studies have reported CO$_2$ and $^{222}$Rn measurement systems in various contexts, including occupational environments, outdoor field investigations, and subterranean karst systems. However, many of these works provide limited methodological transparency, describing only the sensors used while omitting details about the corresponding data logging system \cite{you2012}. Others rely on commercially available data loggers or integrated devices, such as the Radon Scout, AlphaE or AlphaGUARD for radon monitoring, the Vaisala handheld CO$_2$ meter \cite{sainz2022, linan2025}, or the Ahlborn data logger \cite{peyraube2025}. While these systems offer reliable performance, they are typically expensive and lack flexibility for adaptation to site-specific requirements or integration into customized monitoring networks. \\

Since cost factors and flexibility play an important role in making research studies feasible, several self-developed data logger systems have been described in the literature. Many of them are based on microcontrollers, such as PIC or Arduino. Kumar et al. \cite{kumar2010} developed a PIC-based multi-channel data logger with integrated LCD display using serial RS-232 communication for wired data transfer to a host system, but without providing any open-source code. Brown et al. \cite{brown2020} and Levintal et al. \cite{levintal2021} both proposed a low-cost CO$_2$ monitoring system using microcontrollers like Arduino Uno and Adalogger on a printed circuit board, notable for its fully open-source hardware and software design. Kumar et al. \cite{kumar2017} combined an Arduino board with a Raspberry Pi for IoT-based air quality monitoring, where the Pi acted as the major control node; however, the implementation details and code were not shared. These examples highlight the diversity of self-developed logger systems but also the lack of open and adaptable designs suitable for long-term environmental monitoring in complex field conditions.\\

The objective of our CO$_2$ and $^{222}$Rn measurement setup, including a complete description and open-source code of the self-developed Raspberry Pi-based data logger, is to provide a cost-effective, flexible, and user-friendly solution. Compared to simple microcontroller-based systems, the Raspberry Pi offers significant advantages, including advanced networking and communication capabilities, larger storage capacity, and easier programmability, as it functions as a full-fledged computer running Python on a Linux platform. This combination allows for seamless integration of multiple sensors, reliable long-term data acquisition, and remote data access, making it particularly suitable for challenging environments such as underground karst systems.\\

\section{Hardware description}

The hardware system consists of commercial off-the-shelf components: Raspberry Pi units (Raspberry Pi Zero or higher), RS-485-to-USB adapters or RS-485 HATs, power supplies, and standard wiring assembled following a basic wiring diagram. The key distinction from existing sensor logging solutions is the combination of the AtmosPyre Python library for Modbus communication with an Ansible-based deployment system. Configuration requires only editing a YAML inventory file with sensor parameters; the deployment script handles virtual environment setup, dependency installation, systemd service creation, and automatic data logging initialization. The system works with any Modbus sensor, though sensors must first be implemented as AtmosPyre drivers to benefit from single-line Ansible deployment. Currently implemented sensors include the Vaisala GMP252 (CO$_2$) and RadonTech AlphaTRACER ($^{222}$Rn), with additional sensors requiring approximately 100 lines of Python following the provided driver template. Data is written to CSV files with JSON metadata, providing direct compatibility with standard scientific analysis tools.
\newline

\textbf{Key advantages for researchers:}

\begin{itemize}
\item \textbf{Protocol-based universality}: Any RS-485 Modbus sensor can be integrated. Implemented sensors deploy with a single command; new sensors require a Python driver (~100 lines) following documented patterns, after which they gain the same deployment capabilities.

\item \textbf{Deployment speed}: Ansible automation handles configuration, dependency management, and service installation automatically. Networks can be replicated across sites or reconfigured for new experiments in minutes rather than hours.

\item \textbf{Low cost}: Raspberry Pi Zero units start around \euro{}15. No proprietary software licenses are required. 

\item \textbf{Data format}: Timestamped CSV output works directly with Python, R, MATLAB, and spreadsheet software.

\item \textbf{Transparency and extensibility}: Both hardware assembly and software operation are fully visible and modifiable. Researchers can add new sensors following clear templates.

\item \textbf{Modification}: Raspberry Pis can be tailored to specific use cases. For example, they can be used for mobile internet connection via an additional HAT or for mobile power supply using a battery.
\end{itemize}

\section{Design files summary}

\vskip 0.1cm
\tabulinesep=1ex
\begin{tabu} to \linewidth {|X|X|X[1.5,1]|X[1.5,1]|} 
\hline
\textbf{Design filename} & \textbf{File type} & \textbf{Open source license} & \textbf{Location of the file} \\\hline
wiring\_diagram\_usb & pdf & GPL 3.0 or later & \cite{DesignFiles} \\\hline
wiring\_diagram\_usb & svg & GPL 3.0 or later & \cite{DesignFiles} \\\hline
wiring\_diagram\_hat & pdf & GPL 3.0 or later & \cite{DesignFiles} \\\hline
wiring\_diagram\_hat & svg & GPL 3.0 or later & \cite{DesignFiles} \\\hline
\end{tabu}

\vskip 0.3cm
\noindent

There are two exemplary setups presented in this article: a minimal setup using an RS-485 HAT with a Raspberry Pi Zero, and a maximal setup using an RS-485-to-USB converter with a Raspberry Pi 4. The wiring diagrams of these setups are available in SVG and PDF respectively.

\section{Bill of materials summary}

In the following, two setups are showcased. The first is the minimal HAT setup using a Raspberry Pi Zero, while the second is the maximal USB converter setup using a Raspberry Pi 4 for greater flexibility. Standard items such as wires and clamps are not explicitly listed below. Depending on the sensors of interest, a different power supply setup might be more suitable. However, the key components remain as listed below.

\subsection{Minimal HAT setup}
This configuration represents the most cost- and power-effective deployment.
\vskip 0.2cm
\tabulinesep=1ex
\noindent
\begin{tabu} to \linewidth {|X[1.1,1]|X|X[0.6,1]|X[0.8,1]|X|X|X[1.1,1]|}
\hline
\textbf{Designator} & \textbf{Component} & \textbf{Number} & \textbf{Cost per unit - currency} & \textbf{Total cost - currency} & \textbf{Source of materials} & \textbf{Material type} \\\hline
RPI-01 & Raspberry Pi Zero WH & 1 & 16.70 EUR & 16.70 EUR & \href{https://www.reichelt.de/de/de/shop/produkt/raspberry_pi_zero_wh_v_1_1_1_ghz_512_mb_ram_wlan_bt-222531}{reichelt.de} & Semiconductor \\\hline
HAT-01 & Raspberry Pi Zero Shield - RS-485 CAN HAT, MCP2515 & 1 & 12.70 EUR & 12.70 EUR & \href{https://www.reichelt.de/de/de/shop/produkt/raspberry_pi_zero_shield_-_rs485_can_hat_mcp2515-242794}{reichelt.de} & Semiconductor \\\hline
PSU-02 & MeanWell MDR-40-5 DIN-Rail Power Supply, 5V 6A & 1 & 20.47 EUR & 20.47 EUR & \href{https://www.conrad.de/de/p/mean-well-mdr-40-5-hutschienen-netzteil-din-rail-5-v-dc-6-a-30-w-anzahl-ausgaenge-1-x-inhalt-1-st-1293338.html}{conrad.de} & Semiconductor \\\hline
SD-01 & MicroSD Card 8GB Class 10 & 1 & 3.95 EUR & 3.95 EUR & \href{https://www.reichelt.de/de/de/shop/produkt/microsdhc-speicherkarte_8gb_intenso_class_10-126586}{reichelt.de} & Semiconductor \\\hline
\multicolumn{6}{|r|}{\textbf{Total cost (excluding sensors):}} & \textbf{53.82 EUR} \\\hline
\end{tabu}\\
\vskip 0.5cm

\subsection{Maximal USB converter setup}

This configuration provides more flexibility, suitable for laboratory environments or long-term monitoring stations where power consumption is not a primary concern.
\vskip 0.2cm
\tabulinesep=1ex
\noindent
\begin{tabu} to \linewidth {|X[1.1,1]|X|X[0.6,1]|X[0.8,1]|X|X|X[1.1,1]|}
\hline
\textbf{Designator} & \textbf{Component} & \textbf{Number} & \textbf{Cost per unit - currency} & \textbf{Total cost - currency} & \textbf{Source of materials} & \textbf{Material type} \\\hline
RPI-02 & Raspberry Pi 4 Model B (4GB RAM) & 1 & 42.80 EUR & 42.80 EUR & \href{https://www.reichelt.com/de/en/shop/product/raspberry_pi_4_b_4x_1_5_ghz_1_gb_ram_wlan_bt-259874}{reichelt.de} & Semiconductor \\\hline
PSU-01 & Official Raspberry Pi USB-C Power Supply 5.1V 3A & 1 & 7.10 EUR & 7.10 EUR & \href{https://www.reichelt.com/de/en/shop/product/raspberry_pi_-_power_supply_5_1_v_3_0_a_usb_type-c_eu_plug_-260010}{reichelt.de} & Semiconductor \\\hline
CONV-01 & Waveshare Industrial USB to RS-485 Converter (CH343G) & 1 & 8.50 EUR & 8.50 EUR & \href{https://www.berrybase.de/waveshare-industrieller-usb-rs485-konverter-blitzschutz-esd-sicher-bidirektional-ch343g}{berrybase.de} & Semiconductor \\\hline
SD-01 & MicroSD Card 8GB Class 10 & 1 & 3.95 EUR & 3.95 EUR & \href{https://www.reichelt.de/de/de/shop/produkt/microsdhc-speicherkarte_8gb_intenso_class_10-126586}{reichelt.de} & Semiconductor \\\hline
\multicolumn{6}{|r|}{\textbf{Total cost (excluding sensors):}} & \textbf{62.35 EUR} \\\hline
\end{tabu}\\
\vskip 0.2cm
\noindent

\subsection{Housing}
\vskip 0.2cm
\tabulinesep=1ex
\noindent
\begin{tabu} to \linewidth {|X[1.1,1]|X|X[0.6,1]|X[0.8,1]|X|X|X[1.1,1]|}
\hline
\textbf{Designator} & \textbf{Component} & \textbf{Number} & \textbf{Cost per unit - currency} & \textbf{Total cost - currency} & \textbf{Source of materials} & \textbf{Material type} \\\hline
MOUNT-01 & Raspberry Pi DIN rail mounting clip & 1 & 15.04 EUR & 15.04 EUR & \href{https://www.amazon.de/KKSB-Cases-Raspberry-Compatible-Positions/dp/B0CQ5GYL72}{amazon.de} & Polymer \\\hline
RAIL-01 & DIN rail 35mm & 2 & 7.50 EUR & 15.00 EUR & \href{https://www.reichelt.com/de/en/shop/product/mounting_rail_ts_35_7_5_293_mm-119473}{reichelt.de} & Metal \\\hline
BOX-01 & Control cabinet 300 x 400 x 170 mm (IP65) & 1 & 43.55 EUR & 43.55 EUR & \href{https://www.reichelt.de/de/de/shop/produkt/wandgehaeuse_ip65_300_x_400_x_170_mm_grau-262989}{reichelt.de} & Metal \\\hline
\multicolumn{6}{|r|}{\textbf{Total cost:}} & \textbf{73,59 EUR} \\\hline
\end{tabu}\\

\section{Build instructions}

Both the minimal HAT setup and the maximal USB converter setup follow identical assembly principles, differing only in component selection; see the Design Files \cite{DesignFiles}. The core concept involves mounting a Raspberry Pi with RS-485 communication capability alongside power supplies, then wiring sensor connections through screw terminals or directly to the RS-485 interface. Regardless of the setup chosen, when working with RS-485-based sensors and configuring them, there are a few things you need to know: 
\begin{itemize}
    \item The Raspberry Pi's IP address and hostname 
    \item The username you want to use on you Raspberry Pi
    \item The port the RS-485 signal is passed through
    \item RS-485 settings such as slave address, baud rate, parity, stop bits, byte size, and Modbus mode
\end{itemize}

\begin{figure}
    \centering
    \includegraphics[width=0.5\linewidth]{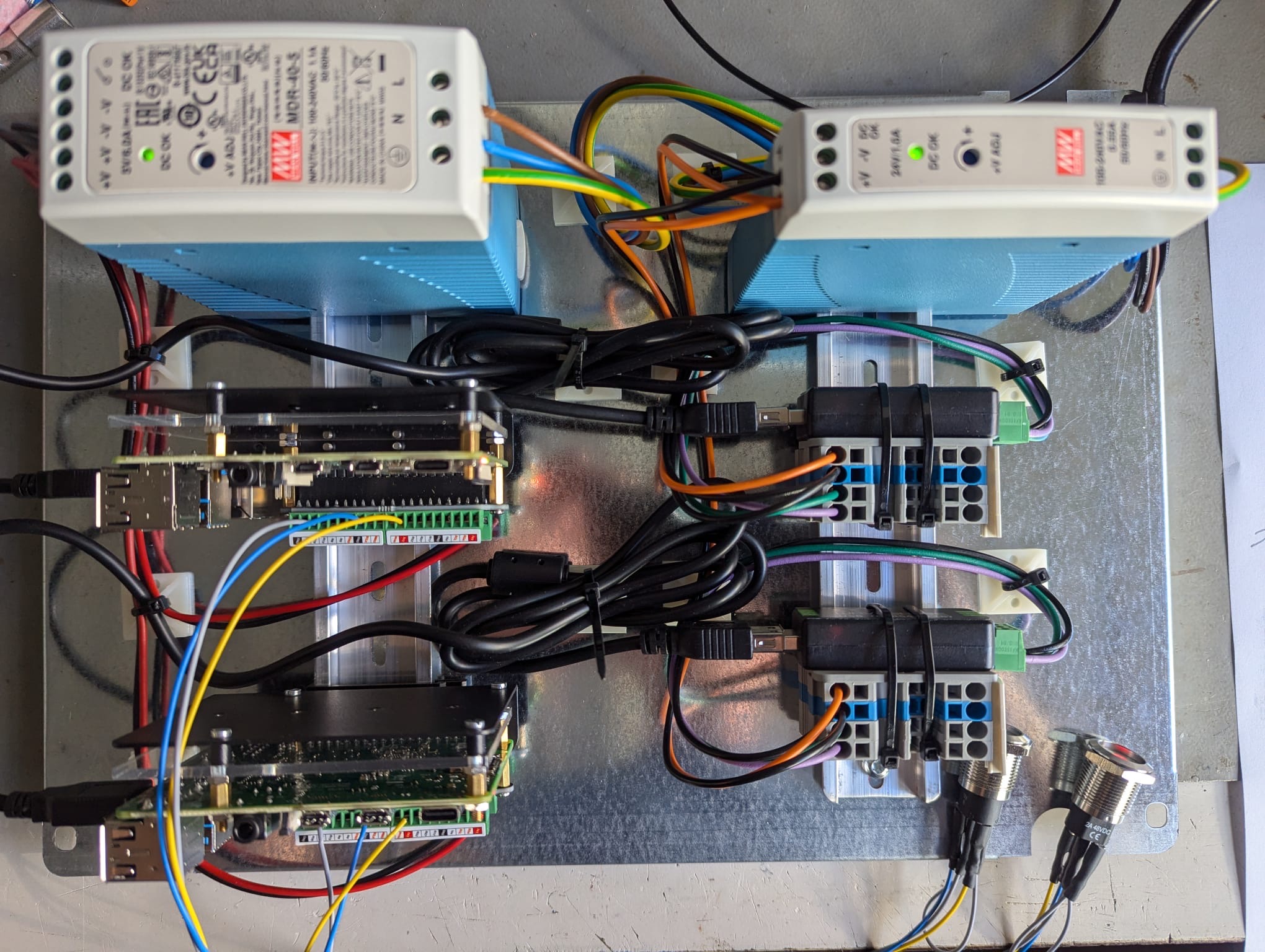}
    \caption{Top view of a setup with redundancy. Two Raspberry Pis simultaneously log CO$_2$. It corresponds to the maximal setup but with a dedicated power supply.}
    \label{fig:top_view}
\end{figure}

\subsection{Safety considerations}

\subsubsection{Electrical safety}

All assembly steps involving mains voltage (230V AC in Europe) must be performed with the system unplugged. If unfamiliar with electrical wiring, consult a qualified electrician for the mains connection. Do not operate the system with exposed mains voltage terminals.

\subsubsection{Sensor power requirements}

Verify sensor voltage requirements before connection. Most environmental sensors operate at 5--24V DC. Incorrect voltage can damage sensors permanently.

\subsubsection{ESD protection} 

Raspberry Pi boards are sensitive to electrostatic discharge.

\subsection{Pre-assembly preparation}

\subsubsection{Raspberry Pi setup}

Install Raspberry Pi OS on the microSD card:
\begin{enumerate}
\item Download Raspberry Pi Imager from \url{https://www.raspberrypi.com/software/} (we recommend using the latest Appimage)
\item Insert microSD card into computer
\item Select a suitable OS; the light versions are recommended.
\item Configure hostname, enable SSH, inject a public SSH key, and set username/password 
\item Take note of the hostname, username and password and make sure you don't lose the SSH key
\item Write image to card
\item Insert card into Raspberry Pi
\item Find the local IP address of the Raspberry Pi
\item Test SSH connection
\begin{verbatim}
ssh username@IPAddress
\end{verbatim}
\end{enumerate}

\subsection{Mechanical assembly}

\subsubsection{Component mounting}

For the minimal HAT setup: Attach the HAT to the Raspberry Pi Zero GPIO header before mounting. Ensure the HAT is firmly seated on all 40 pins.

For the maximal USB converter setup: Plug the USB-to-RS-485 converter into the Raspberry Pi 4 and secure it against bending the port.

\subsection{Electrical connections}
\subsubsection{Raspberry Pi power connection}

Here you have two options depending on your power source. You can either power the Raspberry Pi via the GPIO pins or use a USB port and the official power supply.

Power via GPIO pins:
\begin{enumerate}
\item +V output terminal on 5V power supply (PSU-02) $\rightarrow$ GPIO PIN 4
\item -V (GND) terminal on 5V power supply (PSU-02) $\rightarrow$ GPIO PIN 6
\end{enumerate}

Power via USB:
\begin{enumerate}
\item Plug the official Raspberry Pi USB power supply (PSU-01) directly into the Raspberry Pi. The type of USB will depend on the chosen Raspberry Pi model.
\end{enumerate}

\subsubsection{RS-485 wiring}

RS-485 uses a differential pair (A and B lines) plus ground for communication. Sensors connect via three-wire or four-wire cable depending on the model.

\begin{enumerate}
\item Identify RS-485 terminals on the HAT or USB adapter (typically labeled A, B, GND)
\item Connect sensor wires (read the sensor manual carefully):
\item Sensor A (or D+) $\rightarrow$ HAT/USB adapter A
\item Sensor B (or D-) $\rightarrow$ HAT/USB adapter B
\item For the minimal HAT setup:
\begin{itemize}
\item Sensor GND $\rightarrow$ Raspberry Pi GPIO GND
\end{itemize}
\item For the maximal USB converter setup:
\begin{itemize}
\item Sensor GND $\rightarrow$ USB adapter GND
\end{itemize}
\end{enumerate}

\subsubsection{Sensor power connection}

\begin{enumerate}
\item Verify sensor voltage requirement (typically 5--24V for environmental sensors)
\item Connect sensor power wires:
\begin{itemize}
\item Sensor V+ $\rightarrow$ +V output terminal on sensor power supply
\item Sensor V- or GND $\rightarrow$ -V (GND) terminal on sensor power supply
\end{itemize}
\item Ensure polarity is correct; reversed polarity can damage sensors
\end{enumerate}

\subsubsection{Multiple sensor configurations}

For laboratory setups monitoring multiple sensors simultaneously:
\begin{enumerate}
\item Connect all sensors to the same RS-485 bus (parallel connection of A, B, GND lines). For larger installations, a terminal resistor may be needed.
\item Ensure each sensor has a unique Modbus slave address
\end{enumerate}

\subsubsection{Network connection}

Depending on the network environment, contacting the system administrator may be necessary before proceeding.

Connect to the Raspberry Pi via SSH:
\begin{verbatim}
ssh username@IPAddress
\end{verbatim}

\subsubsection{Deployment of sensors}

% CHANGE: "RS485" -> "RS-485"
Verify that the RS-485 interface is detected:

For the maximal USB converter setup:
\begin{verbatim}
ls /dev/ttyUSB* 
ls /dev/ttyACM*
# Should show /dev/ttyUSB0, /dev/ttyACM0 or similar
\end{verbatim}
Take note of the port for later.

For the minimal HAT setup, the Ansible deployment handles the port automatically to use:
\begin{verbatim}
/dev/ttyAMA0
\end{verbatim}

\subsection{Software deployment}

Software deployment is automated via Ansible.

Basic steps:

\begin{enumerate}

\item Install Ansible on the control computer:
\begin{verbatim}
pip install ansible
\end{verbatim}

\item Clone the deployment repository:
\begin{verbatim}
git clone https://git.iws.uni-stuttgart.de/measurements/senso_pi.git
cd senso_pi
\end{verbatim}
\item Install dependencies:
\begin{verbatim}
    ansible-galaxy install -r requirements.yml
\end{verbatim}
\item Create a custom inventory:
\begin{verbatim}
    cp sensor_inventory.yml my_inventory.yml
\end{verbatim}
\item Modify the inventory for the specific setup:
\begin{itemize}
\item IP address or hostname of the Raspberry Pi
\item The username you used when you set up the Raspberry Pi
\item Sensor type, serial port, Modbus address
\item Measurement tags, logging interval, and output directory
\end{itemize}

\item Full setup and deployment to the Raspberry Pi:
\begin{verbatim}
ansible-playbook -i my_inventory.yml deploy_sensors.yml
\end{verbatim}

\item Redeployment to the Raspberry Pi:
\begin{verbatim}
ansible-playbook -i my_inventory.yml deploy_sensors.yml --tag deploy
\end{verbatim}
\end{enumerate}

The deployment script automatically:
\begin{itemize}
\item Creates a Python virtual environment
\item Installs the AtmosPyre library and dependencies
\item Generates a sensor-specific logging script
\item Configures data output directories
\item Sets up a systemd service (if requested)
\item Starts data logging
\end{itemize}

\subsection{Design alternatives and customization}

\subsubsection{Enclosure selection}

The enclosures shown are weatherproof for outdoor deployment. For indoor laboratory use, simpler plastic project boxes reduce cost. Ensure adequate ventilation if operating in high ambient temperatures.

\section{Operation instructions}

\subsection{Safety considerations}

\subsubsection{Electrical safety}
Do not open the enclosure while connected to mains power. Disconnect power before any maintenance or inspection.

\subsubsection{Environmental limits}
Operation is limited by enclosure IP rating and component specifications.

\subsubsection{Sensor handling}
Follow manufacturer safety guidelines for specific sensors.

\subsection{Normal operation}

Once deployed, the system operates autonomously with no user interaction required. The systemd service ensures automatic startup after power interruption and continuous data logging at configured intervals.

\subsubsection{System status check}

Monitor system health via SSH (where \texttt{[hostname]} corresponds to the hostname configured in the Ansible inventory):

\begin{verbatim}
ssh username@IPAddress
systemctl status sensor-logging-[hostname]
\end{verbatim}

Active status indicates normal operation. Failed status requires log inspection:

\begin{verbatim}
journalctl -u sensor-logging-[hostname] -n 50
\end{verbatim}

\subsubsection{Data retrieval}

Data files are organized by sensor and date in the \texttt{sensor\_logging/data/} directory. Each sensor creates a subdirectory containing daily CSV files with ISO 8601 timestamps.

Copy data to a local machine using SCP:

\begin{verbatim}
scp -r username@IPAddress:~/sensor_logging/data/ ./local_data/
\end{verbatim}

For regular automated backups, use rsync:

\begin{verbatim}
rsync -avz username@IPAddress:~/sensor_logging/data/ ./local_data/
\end{verbatim}

\subsubsection{Stopping and removing the logging service}

To stop the data logging service:

\begin{verbatim}
sudo systemctl stop sensor-logging-[hostname]
\end{verbatim}

To prevent it from restarting after a reboot:

\begin{verbatim}
sudo systemctl disable sensor-logging-[hostname]
\end{verbatim}

To fully remove the service file:

\begin{verbatim}
sudo rm /etc/systemd/system/sensor-logging-[hostname].service
sudo systemctl daemon-reload
\end{verbatim}

\section{Validation and characterization}

The system has been deployed since spring 2025 to measure CO$_2$ and $^{222}$Rn in a karst cave environment characterized by high humidity (near saturation). Over the continuous monitoring period to date, all expected measurements were recorded successfully, with no gaps or corrupted records during system operation.

\begin{figure}[h!]
    \centering
        \centering
        \includegraphics[width=\linewidth]{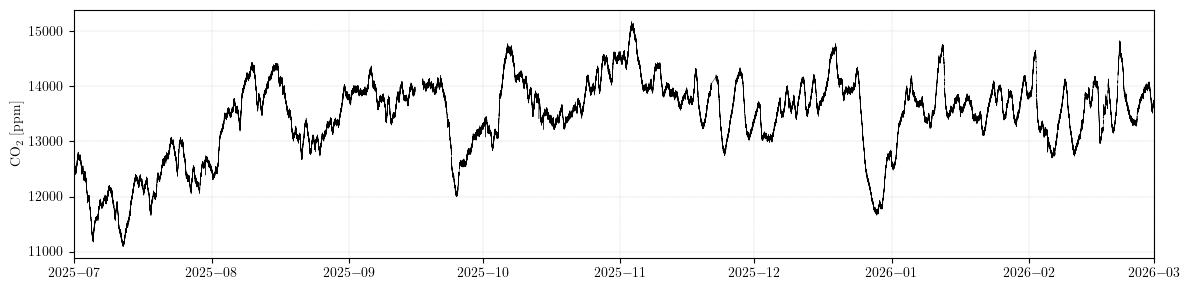}
        \caption{LibrePiLogger setup described in this article}
        \label{fig:newSetup}
\end{figure}

\begin{figure}[h!]
        \centering
        \includegraphics[width=\linewidth]{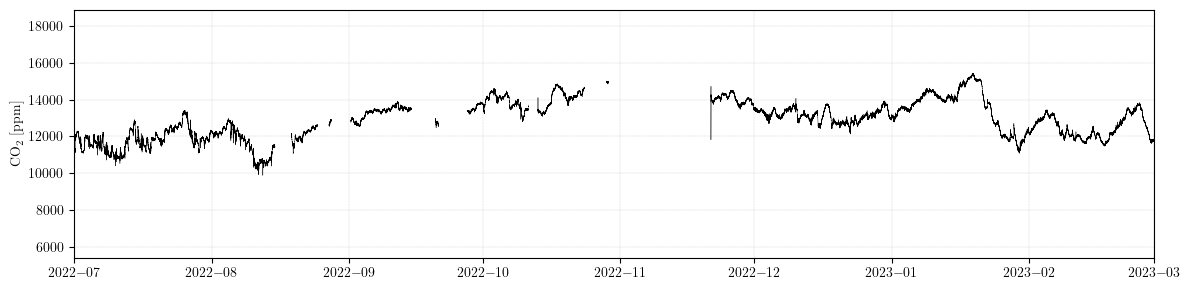}
        \caption{Legacy setup using commercial data loggers}
        \label{fig:oldSetup}
\end{figure}

The proposed setup (\Cref{fig:newSetup}) was validated against a legacy setup using commercial data loggers (\Cref{fig:oldSetup}), operated in parallel on the same site under identical environmental and power conditions. Both systems showed excellent agreement, with indistinguishable temporal trends and only a small systematic offset, remaining within expected sensor uncertainty throughout the parallel operation period.

The system service reliably recovers from power interruptions without user intervention. This behavior has been verified both during unplanned outages in the field (e.g. September 2025) and through repeated testing in the laboratory.
 
Once the Raspberry Pi OS is installed and the hardware is wired, software deployment of new sensors takes approximately 5--10 minutes using the Ansible playbook. This includes inventory configuration, dependency installation, and service activation. Users unfamiliar with the system should expect slightly longer times for initial setup.
 
The Ansible-based deployment also enables remote reconfiguration. During the field deployment, logging parameters such as the measurement interval were updated remotely without physical access to the device. As long as no hardware changes are required, the system can be fully reconfigured over the network.
 
 
\textbf{Capabilities:}
 
\begin{itemize}
\item Continuous autonomous operation over extended periods in challenging subterranean environments (high humidity, limited access)
\item 100\% data completeness during system operation to date
\item Automatic recovery after power interruptions without additional data loss
\item Remote reconfiguration of logging parameters (e.g., measurement interval) without physical access
\item Deployment of a new sensor in approximately 5--10 minutes after hardware assembly
\item Compatible with any RS-485 Modbus sensor; measurement accuracy and precision are determined by the sensor chosen by the user
\item Simultaneous logging of multiple sensors on a single Raspberry Pi via shared RS-485 bus
\end{itemize}
 
\textbf{Limitations:}
 
\begin{itemize}
\item Remote data access, deployment, and reconfiguration require network connectivity (WiFi/Ethernet or mobile network HAT); without network access, data must be retrieved manually via the microSD card
\item Adding a sensor not yet implemented in AtmosPyre requires writing a Python driver (approximately 100 lines), which assumes basic Python knowledge
\item RS-485 bus length is limited
\item The platform does not prescribe a specific power supply or enclosure; users must select these components according to their deployment conditions, as described in the minimal and maximal setup configurations
\end{itemize}

\noindent
\textbf{CRediT author statement}\\

\noindent
\textbf{Leon Keim}: Conceptualization, Software, Validation, Writing - Original Draft, Writing - Review \& Editing. \\
\textbf{Vivien Langhans}: Writing - Original Draft, Writing - Review \& Editing, Data Curation. \\
\textbf{Holger Class}: Supervision, Writing - Original Draft, Writing - Review \& Editing, Funding Acquisition. \\
\textbf{Steffen Hägele}: Conceptualization, Methodology, Resources.\\

\noindent
\textbf{Acknowledgements}\\
\noindent
Funded by Deutsche Forschungsgemeinschaft (DFG, German Research Foundation) under Germany's Excellence Strategy — EXC 2075 – 390740016. We acknowledge the support by the Stuttgart Center for Simulation Science (SimTech). This work was also supported by Project C04 of the Collaborative Research Centre 1313 (SFB 1313, project number 327154368) and internal funds of the University of Stuttgart.
\\

\noindent
\textbf{Declaration of generative AI and AI-assisted technologies in the manuscript preparation process. }

\noindent
During the preparation of this work the authors used Claude (Anthropic) to support code cleaning and refactoring of the AtmosPyre library and deployment scripts in senso-pi, as well as to assist with manuscript writing and language editing. DeepL was additionally used for language and phrasing support. After using these tools, the authors reviewed and edited the content as needed and take full responsibility for the content of the published article.

\noindent

\printbibliography

\end{document}